\documentclass[conference]{IEEEtran}
\IEEEoverridecommandlockouts
\usepackage{cite}
\usepackage{multirow}
\usepackage[pdftex]{graphicx}
\graphicspath{{../pdf/}{../jpeg/}}
\DeclareGraphicsExtensions{.pdf,.jpeg,.png}
\usepackage[cmex10]{amsmath}
\usepackage{amssymb}
\usepackage{algorithmic}
\usepackage{array}
\usepackage{mdwmath}
\usepackage{mdwtab}
\usepackage{eqparbox}
\usepackage{caption}
\usepackage[font=footnotesize]{subfig}
\usepackage{url}
\usepackage{booktabs}
\usepackage{color}
\usepackage{tikz-network}
\usepackage{bm}
\usepackage[linesnumbered,ruled]{algorithm2e}
\usepackage{flushend}

\newcommand\T{\rule{0pt}{2.9ex}}       % Top strut
\newcommand\B{\rule[-1.2ex]{0pt}{0pt}} % Bottom strut

\begin{document}

\title{Temporal Neighbourhood Aggregation: Predicting Future Links in Temporal Graphs via Recurrent Variational Graph Convolutions}

\author{
\IEEEauthorblockN{Stephen Bonner \IEEEauthorrefmark{2},
Amir Atapour-Abarghouei \IEEEauthorrefmark{3}, 
Philip T Jackson \IEEEauthorrefmark{2},
John Brennan \IEEEauthorrefmark{2},
Ibad Kureshi \IEEEauthorrefmark{5}, \\
Georgios Theodoropoulos \IEEEauthorrefmark{4}, 
Andrew Stephen McGough \IEEEauthorrefmark{3} and
Boguslaw Obara \IEEEauthorrefmark{2}}

\IEEEauthorblockA{\IEEEauthorrefmark{2}Department of Computer Science, Durham University, Durham, UK, \\ \{s.a.r.bonner, p.t.g.jackson, j.d.brennan, boguslaw.obara\}@durham.ac.uk} 
\IEEEauthorblockA{\IEEEauthorrefmark{3}School of Computing, Newcastle University, Newcastle, UK,  \ \{amir.atapour-abarghouei, stephen.mcgough\}@newcastle.ac.uk}
\IEEEauthorblockA{\IEEEauthorrefmark{4}School of Computer Science and Engineering, SUSTech, Shenzhen, China, \ georgios@sustec.edu.cn }
\IEEEauthorblockA{\IEEEauthorrefmark{5}Inlecom Systems, Brussels, Belgium,  \ ibad.kureshi@inlecomsystems.com}}

\maketitle

\begin{abstract}

Graphs have become a crucial way to represent large, complex and often temporal datasets across a wide range of scientific disciplines. However, when graphs are used as input to machine learning models, this rich temporal information is frequently disregarded during the learning process, resulting in suboptimal performance on certain temporal inference tasks. To combat this, we introduce Temporal Neighbourhood Aggregation (TNA), a novel vertex representation model architecture designed to capture both topological and temporal information to directly predict future graph states. Our model exploits hierarchical recurrence at different depths within the graph to enable exploration of changes in temporal neighbourhoods, whilst requiring no additional features or labels to be present. The final vertex representations are created using variational sampling and are optimised to directly predict the next graph in the sequence. Our claims are supported by experimental evaluation on both real and synthetic benchmark datasets, where our approach demonstrates superior performance compared to competing methods, outperforming them at predicting new temporal edges by as much as 23\% on real-world datasets, whilst also requiring fewer overall model parameters. 

\end{abstract}


\section{Introduction}
\label{sec:introduction}

Using graphs to represent relationships in large, complex and high-dimensional datasets has become a universal phenomenon across many scientific fields. Encompassing not only computer scientists, interested in social and citation networks \cite{kipf2017semi}, but biologists, studying protein interaction graphs for associations with diseases \cite{yang2014egonet}, chemists, who model molecule properties by treating them as graphs \cite{wu2018moleculenet}, and physicists, who use graphs to model a physical environment \cite{battaglia2016interaction}. 

\begin{figure}[t!]
  \centering

  \begin{tikzpicture}[multilayer=3d]

    \SetLayerDistance{-2.0}
    \Plane[x=-.5,y=-.5, width=3, height=2.5, layer=3]
    \Plane[x=-.5,y=-.5, width=3, height=2.5, layer=2]
    \Plane[x=-.5,y=-.5, width=3, height=2.5, color=green]

    % Layer 3
    \Vertex[x=0., y=0.3, IdAsLabel, Math, layer=1]{v1_3}
    \Vertex[x=1.5, y=1.5, IdAsLabel, Math, layer=1]{v2_3}
    \Vertex[x=1.4, y=0.2, IdAsLabel, Math, layer=1]{v3_3}
    \Vertex[x=0.1, y=1.4, IdAsLabel, Math, layer=1]{v4_3}

    \Edge[bend=10](v1_3)(v2_3)
    \Edge[bend=-30](v1_3)(v3_3)
    \Edge[bend=30](v1_3)(v4_3)

    \Edge[bend=10,color=red](v4_3)(v3_3)
    \Edge[bend=-30,color=red](v2_3)(v4_3)

    \Text[rotation=360, x=0, y=-0.75, layer=1]{$G_T$}

    % Layer 2
    \Vertex[x=0., y=0.3, IdAsLabel, Math, layer=2]{v1_2}
    \Vertex[x=1.5, y=1.5, IdAsLabel, Math, layer=2]{v2_2}
    \Vertex[x=1.4, y=0.2, IdAsLabel, Math, layer=2]{v3_2}
    \Vertex[x=0.1, y=1.4, IdAsLabel, Math, layer=2]{v4_2}

    \Edge[bend=10](v1_2)(v2_2)
    \Edge[bend=-30](v1_2)(v3_2)
    \Edge[bend=30](v1_2)(v4_2)

    \Text[rotation=360, x=0, y=-0.75, layer=2]{$G_2$}

    % Layer 1
    \Vertex[x=0., y=0.3, IdAsLabel, Math, layer=3]{v1_1}
    \Vertex[x=1.5, y=1.5, IdAsLabel, Math, layer=3]{v2_1}
    \Vertex[x=1.4, y=0.2, IdAsLabel, Math, layer=3]{v3_1}
    \Vertex[x=0.1, y=1.4, IdAsLabel, Math, layer=3]{v4_1}

    \Edge[bend=10](v1_1)(v2_1)
    \Edge[bend=-30](v1_1)(v3_1)

    \Edge[Direct, style=dashed, opacity=0.3](v1_1)(v1_3)
    \Edge[Direct, style=dashed, opacity=0.3](v2_1)(v2_3)
    \Edge[Direct, style=dashed, opacity=0.3](v3_1)(v3_3)
    \Edge[Direct, style=dashed, opacity=0.3](v4_1)(v4_3)

    \Text[rotation=360, x=0, y=-0.75, layer=3]{$G_1$}
    
  \end{tikzpicture}
  \caption{The temporal link prediction task is to predict the new edges (red) in the final graph snapshot $G_T$ (green plane) given the previous graphs $G_1$ and $G_2$. }
  \label{fig:temporalLP}
  \vskip -15pt
\end{figure}
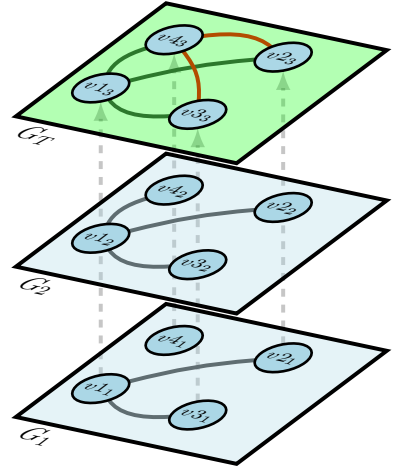

Using graph-based approaches enables complex data analysis, with one of the most universal being the identification of missing links within the graph, which can provide invaluable insight in many real-world scenarios. For example, the recommendation of acquaintances on social networks, new research papers to read or even new links between molecules. However, to date, almost all of the prediction work performed on graphs has been focused on analysis in solely the topological domain, ignoring the rich temporal information inherent in so much of the data represented by graphs (as seen Figure \ref{fig:temporalLP}).  

We formally define a graph $G = (V,E)$ as a finite set of vertices $V$, with a corresponding set of edges $E$. Elements of $E$ are unordered tuples $\{i,j\}$ where $i,j \in V$. Elements in $V$ and $E$ may have labels or certain associated features, although these are not required for this work. In order to perform analysis on graphs, we need a mechanism which converts the formal graph representation into a format which is amenable for machine learning -- graph representation learning.

The field of graph representation learning has received significant attention as a means of analysing large, complex graphs via the use of machine learning. Graph representation learning, comprises a set of techniques that learn latent representations of a graph, which can then be used as the input to machine learning models for downstream prediction tasks \cite{Grover2016}. The majority of graph representation learning techniques have focused upon learning vertex embeddings \cite{Goyal2017review} and reconstructing missing edges \cite{Grover2016}. As such, the goal of graph representation learning is to learn some function $f:V \rightarrow \mathbb{R}^d$ which maps from the set of vertices $V$ to a set of embeddings of the vertices, where $d$ is the required dimensionality. This results in $f$ being a mapping from $G$ to a representation matrix of dimensions $|V| \times d$, i.e. an embedding of size $d$ for each vertex in the graph. However, the majority of graph representation learning approaches to date ignore the temporal aspect of dynamic graphs, resulting in models which perform poorly at predicting future change in a graph. 

This paper introduces a new model, entitled Temporal Neighbourhood Aggregation (TNA), designed to learn vertex representations which capture both topological and temporal change by exploiting the rich information found in large dynamic graphs. To achieve this, we propose a novel model architecture combining graph convolutions with recurrent connections on the resulting vertex level representations to allow for powerful, hierarchical learning at multiple hops of a vertices neighbourhoods. This approach means the model can explore at which neighbourhood depth the most useful temporal information can be learned. Further, we aggregate the temporal neighbourhood using tools from variational inference, resulting in a more robust and stable final representation for each vertex. Our TNA model is trained end to end on temporal graphs represented as time snapshots, where the objective is to directly and accurately predict the next graph in the sequence using the embeddings alone. This results in a model, which unlike many competing approaches, requires no explicitly parameterized decoder model. In summary, our primary contributions are as follows:

\begin{itemize}

  \item \emph{Temporal Neighbourhood Aggregration} - Our proposed model is capable of independently learning the temporal evolutionary patterns within the neighbourhood of a vertex at different depths, resulting in superior performance at predicting future links. Moreover, our approach requires no additional vertex features, labels or random walk procedures as part of its process.
  
  \item \emph{Variational Sampling} - More robust temporal representations and consequently accurate prediction of the next graph in the evolving sequence is made possible by our approach by sampling vertex embeddings using the principals of variational inference. 

  \item \emph{Model Efficacy and Scalability} - Our model contains significantly fewer parameters than competing approaches, as it does not require a parameterized decoder portion. This leads to our model being scalable to larger graphs as a result of its memory efficiency. 
  
\end{itemize}

Our work is supported by extensive experimentation on public benchmark datasets. Further, to aid reproducibility, we open-source all of our PyTorch \cite{paszke2019pytorch} based source-code\footnote{\url{https://github.com/sbonner0/temporal-neighbourhood-aggregation}} and experimentation scripts.
%----------------------------------------------------------------------------------------
%	SECTION - Related Works
%----------------------------------------------------------------------------------------
\section{Related Works}
\label{sec:litreview}

We highlight prior work in the areas of graph representation learning and temporal embeddings.

% Good review of graph conv nets here - https://arxiv.org/pdf/1806.00088.pdf
\subsection{Graph Representation Learning}
\label{ssec:litreview:graph-embedding}

Historically, low dimensional graph representations were created via matrix factorization techniques. Examples of such approaches include Laplican eigenmaps \cite{belkin2002laplacian} and Graph Factorization \cite{ahmed2013distributed}. More recent models, originally used for Natural Language Processing (NLP) tasks, have been adapted to learn graph embeddings. These approaches exploit random walks to create `sentences' which can be used as input to language-inspired models such as DeepWalk \cite{Perozzi2014} and Node2Vec \cite{Grover2016}.

Graph-specific neural network based models have been created, inspired by Convolutional Neural Networks (CNN). Such approaches attempt to create a differential model for learning directly from graph structures. Many Graph CNN approaches operate in the spectral domain of the graph, using eigenvectors derived from the Laplacian matrix of a graph \cite{kipf2017semi}. The Graph Convolutional Network (GCN) approach has proven to be particularly effective \cite{kipf2017semi}. GCN uses a layer-wise propagation rule to aggregate information from the 1-hop neighbourhood of a vertex to create its representation. This layer-wise rule can be stacked $k$ times to aggregate information from $k$ hops away.

The approaches discussed thus far have been supervised, mandating the use of labels. However, graph embedding approaches exist which are based on auto-encoders - a type of neural network trained to reconstruct the input data after initially being projected into a lower dimension \cite{baldi2012autoencoders}. For example, GCNs have been used as the basis of a convolutional auto-encoder model \cite{kipf2016variational}, demonstrating state-of-the-art results for static link prediction.

\subsection{Temporal Graph Embeddings}
\label{ssec:litreview:temporal-embedding}

We argue that the existing approaches for temporal graph embeddings can be split in two categories: Temporal Walk and Adjacency Matrix Factorisation.

\subsubsection{Temporal Walk Approaches}

In an approach entitled STWalk \cite{pandhre2018stwalk}, the authors aim to learn \emph{node trajectories} via the use of random walks which learn representations that consider all the previous time-steps of a temporal graph. In the best performing approach presented, the authors learn two representations for a given vertex simultaneously which are concatenated to create the final temporal embedding. However, the approach is not end to end and requires the user to manually chose how many time steps to consider.

% NetWalk: A Flexible Deep Embedding Approach for Anomaly Detection in Dynamic Networks -----------------
Yu et al.~\cite{Yu2018}, propose NetWalk, which enables anomaly detection in streaming graphs via a vertex-level dynamic graph embedding model. In the approach, a collection of short random walks captured from the graph is passed into an auto-encoder based model to create the vertex representations. 

% Continuous-Time Dynamic Network Embeddings -----------------
Nguyen et al.~\cite{nguyen2018continuous}, propose a model to incorporate temporal information when creating graph embeddings via random walks by capturing individual temporal changes within a graph. They propose a temporal random walk to create the input data, with the approach producing more complex and rich temporal walks via a biasing process.

\subsubsection{Adjacency Matrix Factorisation Approaches}

% DynGEM: Deep Embedding Method for Dynamic Graphs -----------------
Goyal et al.~\cite{Goyal2017}, propose a model for creating dynamic graph embeddings, entitled DynGEM. In this approach, they extend the auto-encoder graph embedding model of Structural Deep Network Embedding (SDNE)~\cite{wang2016structural} to consider dynamic graphs, by using a method similar to Net2net~\cite{Chen2015}, which is designed to transfer knowledge from one neural network to a second.

In a family of approaches entitled Dyngraph2vec*, comprised of DynAE, DynRNN and DynAERNN, Goyal et al.~\cite{goyal2019dyngraph2vec} further extend an SDNE type approach to incorporate temporal information in a variety of ways. The best performing of approaches, DynAERNN, uses a combination of SDNE-like dense auto-encoders, with stacked recurrent layers to learn temporal information when creating vertex embeddings. However, they do not make use of graph convolutions and require a complex decoder model to predict the next graph.

There have been attempts to incorporate temporal aspects into GCNs. However, some \cite{manessi2019dynamic,seo2018structured} focus upon supervised learning, do not explicitly use the models to predict the future graph state and only have a single layer of recurrent connections. More recent approaches, such as GCN-GAN \cite{lei2019gcn} and GC-LSTM \cite{chen2018gc} require large and complex decoder models, meaning they cannot scale to graphs of one-thousand vertices or more on current hardware, whilst also lacking the variational sampling of our approach. In comparison, EvolveGCN \cite{pareja2019evolvegcn} uses recurrent layers to directly evolve the parameters of standard GCN layers which means it does not track vertex neighbourhood evolution explicitly. 

One of the application areas most frequently learning temporal models on graphs is that of traffic modelling. Where approaches like \cite{yao2019revisiting} and \cite{li2017diffusion} combine graph learning with temporal models to predict traffic movement. However, unlike these approaches we focus on creating vertex level embeddings directly optimised to predict future edges and learn change at different hops of a vertices neighbourhood.
%----------------------------------------------------------------------------------------
%	SECTION - Method
%----------------------------------------------------------------------------------------
\section{Methodology}
\label{sec:method}

We briefly outline the proposed approach, relevant background, network architecture and the training procedure. Throughout, we make use of the notation in Table \ref{tab:notation}.

\begin{table}[!t]
  \centering
  \begin{tabular}{c  p{6.5cm}}
    \toprule
   \textbf{Symbol} & \textbf{Definition} \\
   \midrule \midrule
   $G$ & A graph with an associated set of vertices $V$ and corresponding set of edges $E$.\\
   $A$ & The adjacency matrix of graph $G$, a symmetric matrix of size $|V| \times |V|$, where $(a _i,_j)$ is 1 if an edge is present and 0 otherwise. \\
   $\hat A$ & $A$ normalised by its degree matrix $D$ and its identity matrix $I$ such that $\hat A = ({D}^{-\frac{1}{2}} (A+I) {D}^{-\frac{1}{2}})$ \cite{kipf2017semi}.\\
   $X$ & A matrix of features for each $v \in V$, set to the identity $I$ of $A$ for this work. \\
   $H$ & The intermediate vertex representations in GCN and TNA layers. \\
   $Z$ & The final variationally sampled representation matrix for each $v \in V$. \\
   $G^\prime$ & A temporal graph comprised of snapshots $\{G_1, G_2,...,G_T\}$. \\
   $T$ & The number of snapshots in $G^\prime$. \\
   $G_t$ & A graph from $G^\prime$. \\
   $\sigma_s$ & The sigmoid activation function. \\
   $\sigma_r$ & The rectified linear activation function (ReLU). \\
   $\sigma_{lr}$ & The leaky ReLU activation function. \\
   $l$ & A certain layer in the model. \\
   $W^{(l)}_{g}$ & A weight matrix at layer $l$ used in the GCN. \\
   $W^{(l)}_{s}$ & A weight matrix at layer $l$ used in the skip connection. \\
   $W^{(l)}_{\{r,u,h\}}$ & Hidden transform matrices in the GRU. \\
   $U^{(l)}_{\{r,u,h\}}$ & Input transform matrices in the GRU. \\
   $\mathcal{N}(\mu, \sigma)$ & A multi-dimensional Gaussian distribution parametrised by vectors $\mu$ and $\sigma$. \\
   $\Theta$ & A trainable model containing a set of parameters. \B \\
   \bottomrule
  \end{tabular}
  \caption{Definitions and Notations}
  \label{tab:notation}
  \vskip -15pt
\end{table}

\subsection{Motivation}
\label{sec:motivation}

Many of the phenomena that are commonly represented via graph structures are known to evolve over time -- Links between entities form and break in a constantly evolving stream of changes. We thus view graphs as a series of snapshots, with each graph snapshot containing the connections present at that particular moment in time. More formally, we can redefine a graph $G$ to be a temporal graph $G^\prime = \{G_1, G_2,...,G_T\}$, where each graph snapshot $G_{t}$ $ \forall t \in [1,T]$ contains a corresponding vertex set $V_{t}$ and edge set $E_{t}$.

A common and vital task within the field of graph mining is that of future link prediction, where the goal is to accurately predict which vertices within a graph will form a connection in the future \cite{Goyal2017}. Figure \ref{fig:temporalLP} highlights this future link prediction task, where the goal is to predict the new edges, coloured in red, formed in $G_T$, given the previous graphs in the temporal history $G_1$ and $G_2$. Any model designed to accomplish this task must learn the evolution patterns present in edge formation, even though the number of edges changing at each time point is often a small fraction of the total number.

We propose to tackle this by creating temporally-aware graph embeddings, which are explicitly trained to recreate a future time step of the graph. We entitle our approach Temporal Neighbourhood Aggregation (TNA), since to create a better and more meaningful representation for a certain vertex, the model is able to aggregate information about how its neighbourhood has changed in the past to more accurately predict how it will change into the future. More concretely, a temporal graph $G^\prime$ is input to our TNA model $\Theta(G^\prime$) which learns a representation for each vertex in $G_t \in G^\prime$ such that its output can accurately predict the graph $G_{t+1}$. Ideally, we want to create a model $\Theta()$ which can perform this temporal learning using just the sequence of graphs until $G_t$, such that $G_{t+1} = \Theta(G_1,...,G_t)$. TNA is able to accomplish this, requiring no pre-processing steps which could affect the models performance (e.g. random walk procedures), no pre-computed vertex features and no additional labels.

\subsection{Background Technologies}
\label{sec:background}

We first review the background technologies we are employing to make it possible, namely Graph Convolutions \cite{kipf2017semi} and Recurrent Neural Networks \cite{hochreiter1997long, cho2014learning}.

\subsubsection{Graph Convolutions}
\label{sec:gcn}

To perform the graph encoding required to create the initial vertex representations, we utilise the spectral Graph Convolution Networks (GCN) \cite{kipf2017semi}. One can consider a GCN to be a differentiable function for aggregating information from the immediate neighbourhood of vertices \cite{chen2018fastgcn,hamilton2017inductive}. A GCN takes the normalised adjacency matrix $\hat A$ representing a graph $G$, and a matrix of initial vertex level features $X$, and computes a new matrix of vertex level features $H = GCN(\hat A, X)$. $X$ can be initialized with pre-computed vertex features, but it is sufficient to initialize with one-hot feature vectors (in which case $X$ is the identity matrix $I$). A GCN can contain many layers which aggregate the data, where the operation performed at each layer by the GCN \cite{kipf2017semi} is:
\begin{equation}
    \label{eq:GCN}
    GCN^{(l)}(H^{(l)}, \hat A) = \sigma_r (\hat A H^{(l-1)} W^{(l)}_g) \, ,
\end{equation}
where $l$ is the number of the current layer, $W^{(l)}_g$ denotes the weight matrix of that layer, $H^{(l-1)}$ refers to the features computed at the previous layer or is equal to $X$ at $l=0$.

One can consider the GCN function to be aggregating a weighted average of the neighbourhood features for each vertex in the graph. Stacking multiple GCN layers has the effect of increasing the number of hops from which a vertex-level representation can aggregate information -- a three layer GCN will aggregate information from three-hops within the graph to create each representation. 

The original method requires GCN based models to be trained in a supervised learning framework, where the final vertex representation is tuned via labels provided for a specific task -- classification being common \cite{kipf2017semi,hamilton2017inductive}. Extensions to the GCN framework have been made which allow for convolutional auto-encoders for graph datasets \cite{kipf2016variational}.

\subsubsection{Recurrent Neural Networks (RNN)}
\label{sec:rnn}

RNN are neural networks with circular dependencies between neurons. Activations of a recurrent layer are dependent on their own previous activations from a previous forward pass, and therefore form a type of internal state that can store information across time steps. They are frequently used in sequence processing tasks where the response at one time step should depend in some way on previous observations. Long Short-Term Memory (LSTM) \cite{hochreiter1997long} and Gated Recurrent Units (GRU) \cite{cho2014learning} are RNNs with learned gating mechanisms, which mitigate the vanishing gradient problem when back-propagating errors over a sequence of inputs, allowing the model to learn longer-term dependencies. For this work, we employ the GRU cell, as it empirically offers similar performance to an LSTM, but with fewer overall parameters. The GRU computes the output $h_t$, for the input vector $x_t$ at time $t$ in the following manner \cite{cho2014learning}:
\begin{equation}
\begin{aligned}
  \label{eq:GRU}
  u_{t} & =\sigma_s\big(x_{t}U^{(l)}_{u} + h_{t-1}W^{(l)}_{u}\big) \\
  r_{t} & =\sigma_s\big(x_{t}U^{(l)}_{r} + h_{t-1}W^{(l)}_{r}\big) \\
  \tilde{h}_{t} & =\tanh\big(x_{t}U^{(l)}_{h} + (r_{t}\ast h_{t-1})W^{(l)}_{h}\big) \\
  h_{t} & =(1-u_{t})\ast h_{t-1}+u_{t}\ast\tilde{h}_{t},
\end{aligned}
\end{equation}
where $r$ and $u$ are the rest and update gates and $\sigma_s$ and $\tanh$ are the sigmoid and hyperbolic tangent activation functions.

% Model Overview ------------------------------------------------------------------------------------------------------------
\subsection{Model Overview}
\label{sec:model_overview}

We first detail the Temporal Neighbourhood Aggregation blocks which form the primary learning component, before describing the overall model topology and objective function.

\subsubsection{TNA Block}
\label{sec:TNA-Block}

\begin{figure}[t]
  \centering
  \includegraphics[scale=0.55]{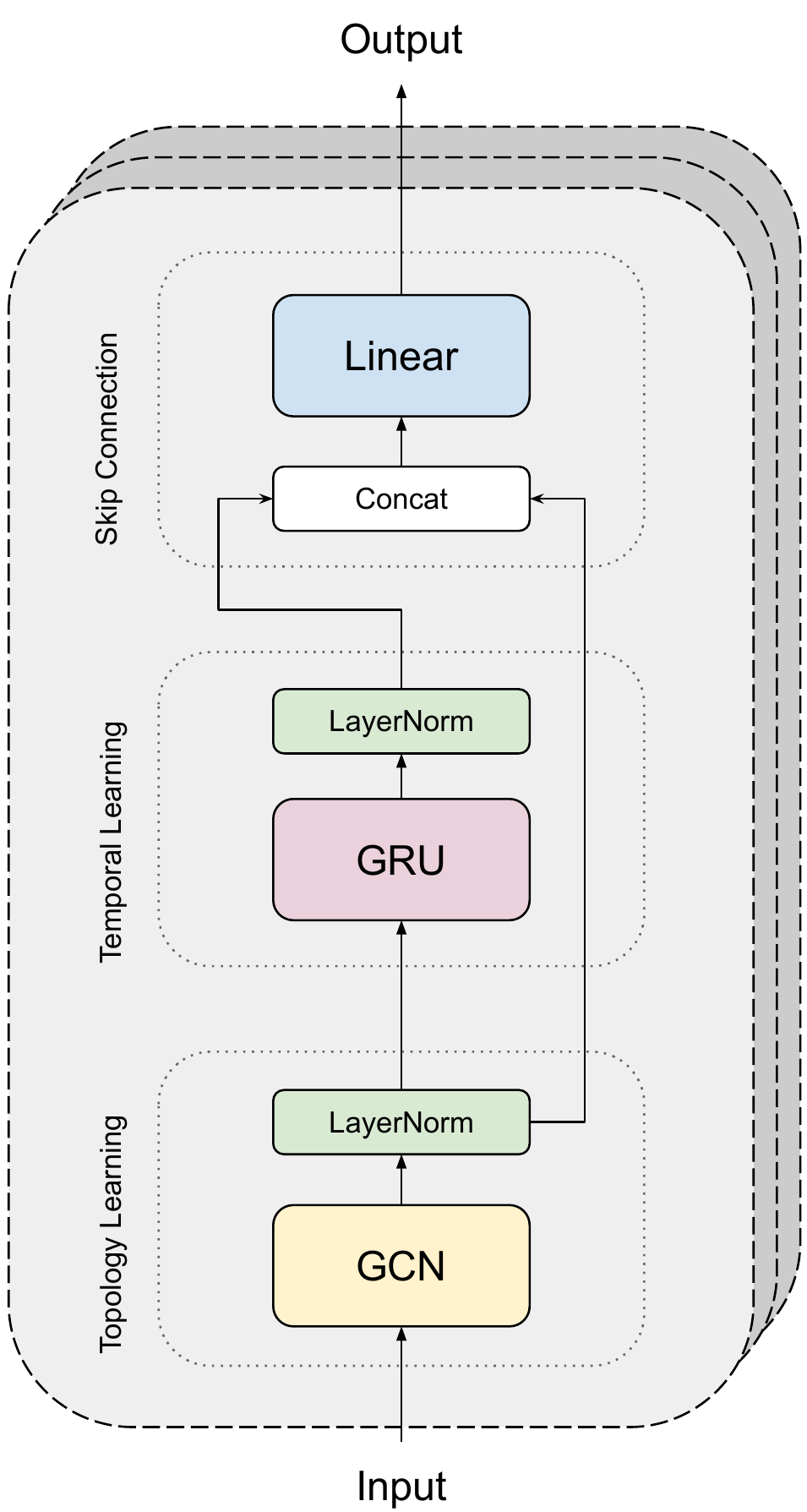}
  \caption{An overview of the Temporal Neighbourhood Aggregation (TNA) block, which comprises a Graph Convolutional Network (GCN) layer with a Gated Recurrent Unit (GRU). The combination of the topological and temporal learning is controlled via the final linear layer.}
  \label{fig:tna-block}
  \vskip -15pt
\end{figure}

\begin{figure*}[t]
  \centering
  \includegraphics[width=\textwidth]{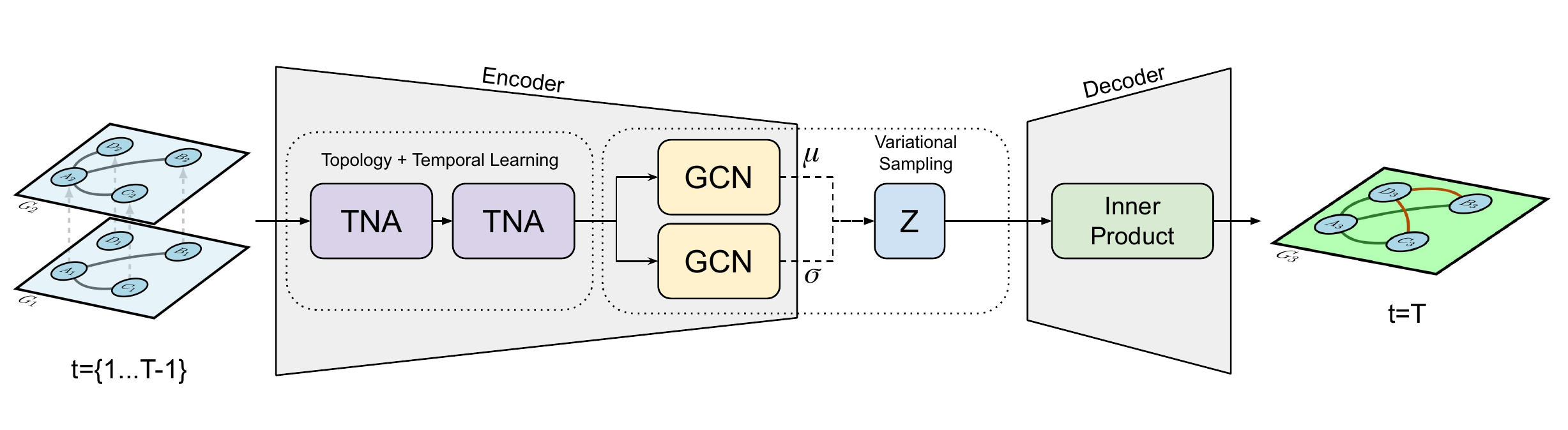}
  \caption{The overall Temporal Neighbourhood Aggregation Model: two stacked TNA blocks learning both topological and temporal information from the first and second hop neighbourhoods of a vertex. An embedding $z_t$ is sampled for each vertex $v_t \in V_t$ using variational inference. The inner product is then used to directly predict the next graph in the sequence.}
  \label{fig:model}
  \vskip -10pt
\end{figure*}

One of the primary components of our model is the TNA block for topological and temporal learning from graphs. The overall structure of the block is illustrated in Figure \ref{fig:tna-block}. It is important to note that all the parameters in the block are shared through time. This allows complex temporal patterns to be learned, as well as allowing for a large reduction in the total number of parameters required by the model. Assuming that the TNA block is the first layer in the model, the flow for vertex $v \in V_t$ can be described as follows:

\begin{itemize}

  \item The input is passed through the GCN layer, as detailed in Equation \ref{eq:GCN}, which will learn to aggregate information for $v$ from its one-hop neighbourhood to create its representation at this point in the block - $h_{t}^{GCN}$. This is then normalised using Layer Norm \cite{ba2016layer}, which will ensure that the representation for each vertex is of a similar scale, this has been shown to improve the training stability and convergence rate of deep models \cite{ba2016layer}. 

  \item  This normalised representation is then passed into a GRU cell a row at a time, as detailed in Equation \ref{eq:GRU},  where the output of the cell will be a function of the current input as well as all the previous inputs. Meaning that the cell can learn how much of the previous neighbourhood representation to use when creating the new representation for a given vertex $h_{t}^{GRU}$. This is then passed through a second Layer Norm unit to ensure a normalised output.

  \item Finally, the $h_{t}^{GCN}$ and $h_{t}^{GRU}$ representations are concatenated together, before being passed through a linear layer and a leaky ReLU activation function to create the final representation for the vertex $h_{t}^{TNA}$. Inspired by residual connections often used in computer vision networks \cite{he2016deep}, this enables the model to learn the optimum mix of topological and temporal information. 

\end{itemize}
The layer-wise propagation rule of the TNA block at depth $l$ can thus be summarised as follows for the entire graph $G_t \in G^\prime$ with normalised adjacency matrix $\hat A_t$:
\begin{equation}
  \begin{aligned}
    \label{eq:TNA_dfdf}
    H_t^{GCN} &= GCN(\hat A_t, H^{(l-1)}_t) \\
    H_t^{GRU} &= GRU(H^{GCN}_t, H_{t-1}^{GRU})  \\
    H_t^{TNA^{(l)}} &= \sigma_{lr}\big(W_{s}^{(l)}\textrm{Concat}(H_t^{GCN}, H_t^{GRU})\big) \\
    TNA(\hat A_t, H^{(l)}_t ) &= H^{(l)}_t = H_t^{TNA^{(l)}}\\
  \end{aligned}
  \end{equation}
where $W_{s}^{(l)}$ represents the weight matrix used to mix the topological and temporal representations, and $\sigma_{lr}$ is the leaky ReLU activation function with a negative slope of 0.01.

\subsubsection{Overall Model Architecture}

As with normal GCN layers, TNA blocks can be stacked to aggregate information from greater depth within a graph, with each additional block adding one extra hop from which information can be aggregated for a certain vertex. However, as our TNA blocks are recurrent, information can also be aggregated from how connectivity within these hops has evolved over time, instead of just their present state. After extensive ablation studies (detailed in Section \ref{sec:ablation}), we use the final configuration of the model detailed in Figure \ref{fig:model}. Our model contains two stacked TNA blocks, to learn information from two hops within the temporal neighbourhood. This is then passed to two independent GCN layers which perform a final aggregation of this temporal representation. From these two layers, the final representation matrix $Z_t$ is sampled using techniques from variational inference, specifically the reparametrisation trick \cite{kingma2013auto}. 

\emph{Variational Sampling -} To create the final representation matrix $Z_t \in \mathbb{R}^{|V_t| \times d}$, the output from the two GCN layers $GCN_{\mu}$ and $GCN_{\sigma}$ are used to parametrise a unit Gaussian distribution $\mathcal{N}$, from which $Z_t$ is then sampled, rather than being explicitly drawn. This is the same concept used in Variational Auto-Encoders \cite{kingma2013auto}, and has previously been demonstrated to work well for creating more robust and meaningful vertex level representations \cite{bonner2018temporal, kipf2016variational}. Our inference model used to create the vertex representations of graph $G_t$, with adjacency matrix $A_t$ and identity matrix of $A_t$, $X_t$, can thus be described as :
\begin{equation}
    \label{eq:TNA-inf}
    q(Z_t|X_t, A_t) = \prod_{v=1}^{|V_t|} \mathcal{N}(z_v | GCN{\mu_v}, \, diag(GCN{\sigma^{2}_v})) ,
\end{equation}
where $q$ is our approximation of the true and intractable distribution we are interested in capturing --  $p(A_{t+1}|Z_t)$. Here, both $GCN_{\mu}$ and $GCN_{\sigma}$ take input from two stacked TNA layers as detailed in Figure \ref{fig:model}. 

\emph{Generative Model -} To decode the information contained within $Z_t$, a generative model is created to explicitly predict the new edges appearing in the next graph in the sequence. Here, the inner-product between the latent representation is used to directly predict $A_{t+1}$: 
\begin{equation}
  \label{eq:TNA-gen}
  p(A_{t+1}|Z_t) = \prod_{i=1}^{|V|} \prod_{j=1}^{|V|} p(A_{t+1_{i, j}}|\sigma_s(z_i z_j^\mathsf{T})),
\end{equation}
where $A_{t+1_{i, j}}$ represents elements from $A_{t+1}$ and $z$ refers to the rows of each vertex taken from $Z_t$.

This generative model is one of the key advantages of our approach, as it means that we have zero learnable parameters in the decoder portion of the model. This is in contrast to many competing approaches, which often require as many parameters as in the encoder to create a decoder with the desired functionality \cite{goyal2019dyngraph2vec}. This results in our approach being able to scale to significantly larger graphs, with longer histories than some of the competing approaches, whilst also being less prone to over-fitting to none-changing edges. 

\subsection{Objective Function}
\label{sec:objective-function}

To train the TNA model, and as is common for variational methods \cite{kingma2013auto, kipf2016variational}, we directly optimise the lower bound $\mathcal{L}$ with regards to the model parameters: 
\begin{equation}
  \begin{gathered}
    \label{eq:GVAE-learn}
    \mathcal{L} = \mathbb{E}_{q(Z_t|X_t, A_t)} \Big[ \text{log} p(A_{t+1}|Z_t) \Big] - \\ 
    KL(q(Z_t| A_t, X_t) || p(Z_t)),
  \end{gathered}
\end{equation}
where $KL()$ is the Kullback-Leibler distance between $p$ and $q$. We use a Gaussian prior as the distribution for $p(Z_t)$. 

In addition, we apply $L_2$ regularization to our model parameters to help with over-fitting, which is defined as:
\begin{equation}
  \mathcal{L}_{reg} = \lambda \sum\limits_{i=1}^{|\Theta|} \Theta_i^{2} ,  
\end{equation}
where $\lambda $ is a scaling factor, set to $10^{-5}$. Consequently, the final objective function for our model is:
\begin{equation}
  \mathcal{L}_{final} = \mathcal{L} + \mathcal{L}_{reg}. 
\end{equation}
\subsection{Model Parameters and Training Procedure}
\label{sec:model_parm}

After initial grid-searches, we empirically found two layers of Temporal Neighbourhood Aggregation, followed by variational sampling, to yield the optimal performance, with the first layer comprising 32 filters, whilst the second having 16 filters. For training the model, we empirically found using full-batch gradient descent with the RMSProp algorithm, a learning rate of 0.001 and 200 epochs to give the best results. Our model has been implemented in PyTorch \cite{paszke2019pytorch}.
%----------------------------------------------------------------------------------------
%	SECTION - Experimental Setup
%----------------------------------------------------------------------------------------
\section{Experimental Setup}
\label{sec:experiment_setup}

We detail the setup of our experimental evaluation, as well as the baseline approaches and the datasets we use.

\subsection{Evaluation Overview and Methodology}
\label{sec:testingmeth}

As the primary goal is to create vertex representations which are better at encoding temporal change, we will be using the task of future link prediction as our primary objective. More formally, we are trying to maximise the probability of $\mathcal{P}(G_t| G_1...G_{t-1})$. In the context of machine learning, this can be defined as training a model from a temporal $G^\prime$ using $ G_1...G_{t-1}$ such that it can predict the new edges in $G_t$, $E_t \setminus E_{t-1}$. The full training and evaluation process is detailed in Algorithm \ref{algo:training_procedure}. Many recent methods attempt to solve this problem via vertex embedding similarity -- i.e. vertices with more similar embeddings, according to some metric, are more likely to be connected via an edge \cite{Grover2016, Perozzi2014, kipf2016variational}. 

Graph edges are predicted as follows: given the learned vertex embeddings, the future adjacency matrix is reconstructed via the dot product of the embedding matrix $A_{t+1}^{\prime} = \sigma(Z_tZ_t^\mathsf{T})$. This reconstructed adjacency matrix is compared with the true graph to assess how well the embedding is able to reconstruct the future graph.

\begin{algorithm}[h]

  \SetKwInOut{Input}{Input}
  \SetKwInOut{Output}{Output}
  
   \Input{The temporal graph $G^\prime = \{G_1, G_2,...,G_T\}$}
  
   \Output{Mean AUC and AP scores for predicting new edges for each graph in $G^\prime$} 
   \For{all $G_t \in G^\prime$ where $t \ge 3$}{
    Load and pre-process the graphs $G_1, G_2,...,G_T$ \\
    Create new model $\Theta_i$ (as shown in Figure \ref{fig:model})\\
    Train $\Theta_i$ on sequence $G_1, G_2,...,G_{t-1}$, where each graph is the input and used to predict the following one \\
    Predict new edges in $G_t$ using $\Theta_i(G_{t-1})$: $E_t \setminus E_{t-1}$ \\
    Store AUC and AP values \\}

    \Return{Mean AUC and AP values over $G^\prime$}

   \caption{New edge predicition procedure}
   \label{algo:training_procedure}   
  \end{algorithm}

\subsection{Performance Metrics}
\label{sec:metrics}

As one can consider the task of link prediction to be a binary classification problem (an edge can only be present or not), we make use of two standard binary classification metrics:

  \begin{itemize}
    \item \emph{Area Under the Receiver Operating Characteristic Curve (AUC)} -- The ratio between the True Positive Rate (TPR) and False Positive Rate (FPR) measured at various classification thresholds. 
    \item \emph{Mean Average Precision (AP)} -- Across the set of test edges: $AP = \frac{ TP } {TP+FP },$ where $TP$ denotes the number of true positives the model predicts, and $FP$ denotes the number of false positives.
  \end{itemize}
For both of the chosen metrics, a larger value indicates more correctly predicted edges. 

\subsection{Datasets}
\label{sec:datasets}

When performing our experimental evaluation, we employ the empirical datasets presented in Table \ref{tab:datasets}. The graphs represent a range of domains, sizes and temporal complexities. 

\begin{table*}[!t]

    \centering
    \begin{tabular}{c c c c c c c c c}
    \toprule
    \textbf{Dataset} & $|V|$ & $|E|$ & \textbf{First Edge} & \textbf{Last Edge} & \textbf{Num Snapshots} & \textbf{\# New Edges} & \textbf{Reference}\T\B \\
    \midrule \midrule
    Bitcoin-Alpha (Bitcoina) & 3,783 & 24,186  & 08/09/2010 & 22/01/2016 & 62 & 227 & \cite{snapnets}\T \\
    Wiki-Vote (Wiki) & 7,115 & 103,689  & 28/02/2005 & 06/01/2008 & 34 & 2963 & \cite{snapnets} \\
    UC Irvine Messages (UCI) & 1,899 & 20,296 & 15/04/2004 & 25/08/2004 & 27 & 513 & \cite{kunegis2013konect}\B \\
    \bottomrule
    \end{tabular}
    \caption{Empirical graph datasets, where \# New Edges is the average number of new edges added between time points.}
    \label{tab:datasets}
    \vskip -15pt
  \end{table*}

\emph{Bitcoin-Alpha (Bitcoina) -} Representing a trust network within a platform entitled Bitcoin Alpha, where edges are formed as users interact and rate each others reputation. The graph covers a range of edges formed between 8th October 2010 and 22nd January 2016, which we partition into 62 monthly snapshots. The task of new edge prediction is thus analogous to predicting if two users are going to interact within the next month. 

\emph{Wiki-Vote (Wiki) -} Representing a vote of escalating user privileges between users and administrators on the Wikipedia website. The graph covers a range of edges formed between 28th March 2004 and 6th January 2008, which we partition into 34 monthly snapshots. The task of new edge prediction within this data is analogous to predicting if two users are going to vote for each other within the next week. 

\emph{UCI-Messages (UCI) -} Representing private messages sent between users on the University of California Irvine social network platform. The graph covers a range of edges formed between 15th April 2004 and 25th October 2004, which we partition into 27 weekly snapshots. The task of new edge prediction would represent the likelihood that two users will exchange messages with each other over the next week.

\subsubsection{Synthetic Datasets}

In addition, we use two synthetic datasets: a Stochastic Block Model (SBM) graph and a randomly perturbed version of the Cora dataset (R-Cora).

\emph{SBM -} A random graph of 3,000 vertices, which evolves over 30 time points using the SBM algorithm \cite{karrer2011stochastic}. The graph contains 3 communities and at each time point, 20 vertices will evolve by switching from one community to another. 

\emph{R-Cora -} To create this synthetic dataset, we take the original Cora dataset representing a citation network, and perturb the graph using the random rewire method \cite{bonner2016deep, bonner2016efficient}. The rewiring process alters a given source graph's degree distribution by randomly altering the source and target of a set number of edges. During this rewiring process, it is not guaranteed that the source or target of the edge will be altered, which indeed is not always possible due to the topology of the graph. Also, the rewiring process does not change the total number of edges or vertices within the graph. We employ \emph{Erd\H{o}s} rewiring, i.e. the resulting topology of the graph begins to resemble a Erd\H{o}s-R\'{e}nyi graph, where the edges are uniformly distributed between vertices.

\subsection{Baseline Approaches}
\label{sec:baselines}

We compare our approach against a variety of state-of-the-art graph representation learning techniques, both static and dynamic. We choose the baselines which compare most directly with our proposed approach, meaning we opt for comparators which take advantage of deep neural networks to create vertex embeddings.

\begin{itemize}

  \item \emph{GAE}\cite{kipf2016variational}: A non-probabilistic Graph Convolutional Auto-encoder (GAE), where the model is trained on $G_{t-1}$ and then directly predicts new edges in $G_t$. 
  \item \emph{GVAE}\cite{kipf2016variational}: A Graph Variational Convolutional Auto-encoder (GVAE), trained in the same manner as the GAE.  
  \item \emph{TO-GAE}\cite{bonner2018temporal}: A GAE model training procedure which enables temporal offset reconstruction, where the model is trained on $G_{t-2}$ to predict $G_{t-1}$. $G_{t-1}$ is subsequently used as input and the ability to predict $G_t$ is measured.
  \item \emph{TO-GVAE}\cite{bonner2018temporal}: A GVAE model trained using the temporal offset reconstruction method.
  \item \emph{DynAE}\cite{goyal2019dyngraph2vec}: A non-convolutional graph embedding model, similar to SDNE \cite{wang2016structural}, extended to temporal graphs by concatenating the rows of the past graphs together before being passed into the model.
  \item \emph{DynRNN}\cite{goyal2019dyngraph2vec}: A non-convolutional graph embedding model, where stacked LSTM units are used to encode the temporal graph directly. The approach also requires a decoder model, also comprised of stacked LSTM units, to reconstruct the next graph from the embedding.
  \item \emph{DynAERNN}\cite{goyal2019dyngraph2vec}\footnote{For the Dyn* family of algorithms, we use the implementations as provided by the authors as part of their DynamicGEM package \cite{goyal2018dynamicgem}.}: A combination of the previous two models, where a dense auto-encoder is used to learn a compressed representation which is passed to stacked LSTM units for temporal learning. It requires a large decoder, with both dense and LSTM layers, to predict the next graph. The E-LSTM-D approach \cite{chen2019lstm} is also extremely similar to this model.
  \item \emph{D-GCN:}\cite{manessi2019dynamic, seo2018structured}: A dynamic GCN, similar to approaches proposed in \cite{manessi2019dynamic} and \cite{seo2018structured}. Here, three stacked GCN layers are used to capture structural information with an LSTM unit used to learn temporal information and produce the final embeddings. To directly predict the next graph, we use an inner-product decoder on the embedding matrix.
  
\end{itemize}

We attempted to compare with GCN-GAN \cite{lei2019gcn} and GC-LSTM \cite{chen2018gc}, but we were unable to get them to scale to the size of graphs we are using for our experimentation.

\subsection{Experimental Environment}

Experimentation was performed on a system with 2 * NVIDIA Titan Xp GPUs, 2.3GHz Intel Xeon E5-2650 v3, 64GB RAM, with Ubuntu Server 18.04 LTS, Python 3.7, CUDA 10.1, CuDNN v7.4 and PyTorch 1.1.

\section{Results}
\label{sec:results}
\begin{table*}[h!]
  \centering
  \resizebox{\textwidth}{!}{
  \begin{tabular}{l l  c c c | c  c c c }
  \toprule
  \textbf{Dataset}  & \textbf{Approach} & \multicolumn{3}{c}{\textbf{AUC}} & \multicolumn{3}{c}{\textbf{AP}}& $|\Theta|$\T\B \\
  \midrule \midrule
  & & 25\%  & 50\%  & 100\% & 25\%  & 50\%  & 100\%\B \\
  \cline{3-5} \cline{6-8}

\multirow{9}{*}{Bitcoina} & GAE  & $0.466\pm0.025$  & $0.497\pm0.042$  & $0.531\pm0.127$  & $0.613\pm0.031$  & $0.643\pm0.042$ & $0.681\pm0.093$ & $121K$\T \\
                          & GVAE  & $0.577\pm0.048$  & $0.602\pm0.046$  & $0.620\pm0.083$  & $0.634\pm0.043$  & $0.654\pm0.040$ & $0.670\pm0.068$ & $122K$ \\ 
                          & TO-GAE  & $0.551\pm0.053$  & $0.566\pm0.053$  & $0.576\pm0.124$  & $0.694\pm0.038$  & $0.701\pm0.038$ & $0.715\pm0.085$ & $120K$  \\
                          & TO-GVAE  & $0.598\pm0.048$  & $0.620\pm0.045$  & $0.631\pm0.081$  & $0.646\pm0.044$  & $0.665\pm0.040$ & $0.631\pm0.081$ & $122K$  \\
                          & DynAE  & $0.281\pm0.080$  & $0.247\pm0.065$  & $0.209\pm0.071$  & $0.435\pm0.012$  & $0.442\pm0.012$ & $0.439\pm0.023$ & $4.16M$  \\
                          & DynRNN  & $0.181\pm0.081$  & $0.170\pm0.059$  & $0.155\pm0.066$  & $0.388\pm0.014$  & $0.388\pm0.011$ & $0.393\pm0.022$ & $69.9M$  \\
                          & DynAERNN  & $0.093\pm0.090$  & $0.071\pm0.066$  & $0.048\pm0.054$  & $0.326\pm0.022$  & $0.320\pm0.016$ & $0.318\pm0.012$ & $6.98M$  \\
                          & D-GCN & $0.622\pm0.084$  & $0.572\pm0.080$  & $0.519\pm0.144$  & $0.697\pm0.058$  & $0.661\pm0.058$ & $0.623\pm0.107$ & $125K$  \\

                          \cline{2-5} \cline{6-9}
                         & \textbf{TNA}  & $\bm{0.665\pm0.067}$  & $\bm{0.698\pm0.075}$  & $\bm{0.775\pm0.110}$  & $\bm{0.762\pm0.048}$  & $\bm{0.792\pm0.054}$ & $\bm{0.849\pm0.079}$ & $133K$\T  \\
\midrule
\multirow{9}{*}{UCI} & GAE  & $0.561\pm0.075$  & $0.600\pm0.075$  & $0.606\pm0.092$  & $0.661\pm0.066$  & $0.688\pm0.060$ & $0.689\pm0.079$ & $61K$  \\ 
                     & GVAE  & $0.571\pm0.079$  & $0.606\pm0.074$  & $0.619\pm0.065$  & $0.585\pm0.059$  & $0.621\pm0.063$ & $0.625\pm0.060$ & $62K$  \\
                    & TO-GAE  & $0.601\pm0.059$  & $0.633\pm0.061$  & $0.625\pm0.087$  & $0.682\pm0.053$  & $0.705\pm0.050$ & $0.699\pm0.076$ & $61K$  \\
                    & TO-GVAE  & $0.582\pm0.072$  & $0.614\pm0.069$  & $0.624\pm0.062$  & $0.590\pm0.057$  & $0.624\pm0.062$ & $0.627\pm0.060$ & $62K$  \\
                    & DynAE  & $0.234\pm0.066$  & $0.168\pm0.076$  & $0.128\pm0.067$  & $0.436\pm0.019$  & $0.435\pm0.021$ & $0.433\pm0.017$ & $2.28M$  \\
                    & DynRNN  & $0.161\pm0.019$  & $0.176\pm0.024$  & $0.159\pm0.048$  & $0.365\pm0.016$  & $0.370\pm0.016$ & $0.369\pm0.029$ & $21.8M$  \\
                    & DynAERNN  & $0.033\pm0.032$  & $0.021\pm0.025$  & $0.013\pm0.019$  & $0.314\pm0.005$  & $0.312\pm0.004$ & $0.312\pm0.003$ & $4.15M$  \\
                    & D-GCN  & $0.508\pm0.041$  & $0.555\pm0.071$  & $0.565\pm0.068$  & $0.605\pm0.045$  & $0.653\pm0.066$ & $0.656\pm0.072$ & $64K$  \\    

                    \cline{2-5} \cline{6-9}
                    & \textbf{TNA}  & $\bm{0.694\pm0.077}$  & $\bm{0.749\pm0.073}$  & $\bm{0.764\pm0.071}$  & $\bm{0.702\pm0.073}$  & $\bm{0.763\pm0.075}$ & $\bm{0.783\pm0.067}$ & $72K$\T  \\
                  \midrule

\multirow{8}{*}{Wiki} & GAE  & $0.491\pm0.035$  & $0.487\pm0.038$  & $0.502\pm0.040$  & $0.642\pm0.029$  & $0.621\pm0.033$ & $0.617\pm0.032$ & $228K$  \\  
                      & GVAE  & $0.580\pm0.024$  & $0.573\pm0.018$  & $0.563\pm0.024$  & $0.598\pm0.032$  & $0.589\pm0.025$ & $0.572\pm0.029$ & $229K$  \\
                      & TO-GAE  & $0.537\pm0.052$  & $0.556\pm0.049$  & $0.552\pm0.048$  & $0.700\pm0.032$  & $0.697\pm0.027$ & $0.668\pm0.044$ & $228K$  \\
                      & TO-GVAE  & $0.599\pm0.028$  & $0.595\pm0.021$  & $0.579\pm0.029$  & $0.613\pm0.036$  & $0.604\pm0.029$ & $0.583\pm0.034$ & $229K$  \\
                      & DynAE  & $0.354\pm0.034$  & $0.325\pm0.041$  & $0.244\pm0.089$  & $0.448\pm0.009$  & $0.463\pm0.016$ & $0.467\pm0.013$ & $7.5M$  \\
                      & DynAERNN  & $0.183\pm0.024$  & $0.179\pm0.026$  & $0.127\pm0.056$  & $0.342\pm0.005$  & $0.341\pm0.006$ & $0.329\pm0.012$ & $11.9M$  \\
                      & D-GCN  & $0.628\pm0.160$  & $0.591\pm0.115$  & $0.563\pm0.087$  & $0.745\pm0.104$  & $0.686\pm0.094$ & $0.629\pm0.089$ & $231K$  \\ 
                      
                      \cline{2-5} \cline{6-9}
                      & \textbf{TNA} & $\bm{0.674\pm0.034}$  & $\bm{0.644\pm0.044}$  & $\bm{0.634\pm0.050}$  & $\bm{0.759\pm0.025}$  & $\bm{0.740\pm0.032}$ & $\bm{0.736\pm0.039}$ & $239K$\T  \\

  \bottomrule
  \end{tabular}}
  \caption{Next graph prediction results presented as mean values with standard deviation when predicting at various percentages of the length of the time-sequence. A bold value indicates the highest score for that metric. The number of parameters required by each model for the specific datasets are also included.}
  \label{tab:lp}
  \vskip -10pt
\end{table*}

We evaluate our TNA approach using comparisons against state-of-the-art approaches and ablation studies using well-established datasets (Section \ref{sec:datasets}).

\subsection{Ablation Study}
\label{sec:ablation}

One of the major contributions of our work is highlighting how each component of our TNA model is crucial in producing good temporal embeddings. To highlight this, Table \ref{tab:ablation} shows how adding components of the model sequentially affects the performance of predicting new edges in the final graph of the Bitcoina dataset. It is important to note that adding temporal information from both the first and second hop neighbourhood (Model TTV) lifts both AUC and AP scores by approximately 10\% versus just first hop temporal information (Model TGV). This supports our hypothesis that a vertex requires temporal information from more than just its first-order neighbourhood in order to predict future edges. The ablation study also demonstrates that, with a modest increase in the number of parameters, the temporal models are able to exploit the rich information available in the graph's past evolution to much more accurately predict future edges. 

\begin{table}[h!]
  \centering
  \begin{tabular}{l c c c }
  \toprule
  \textbf{Approach}  & \textbf{AUC} & \textbf{AP} & $|\Theta|$  \T\B \\
    \midrule
    \midrule

    GGG            &  $0.574$ & $0.747$ & $121K$  \\
    GGV          &  $0.721$ & $0.705$ & $122K$  \\
    TGV       &  $0.772$ & $0.809$ & $130K$  \\
    TTV       &  $0.863$ & $0.916$ & $132K$  \\
    TTV/LN    &  $0.927$ & $0.932$ & $132K$\B \\ \hline
    TTV/LN/SC (TNA) &  $\mathbf{0.977}$ & $\mathbf{0.976}$ & $133K$\T\B  \\
    
  \bottomrule
  \end{tabular}
  \caption{Ablation study results on the Bitcoina dataset. G is a GCN layer, V is a varitonal sampling layer, T is a GCN + GRU layer, LN is Layer Norm and SC is a skip-connection. $|\Theta|$ is the total number of learnable parameters in the model.}
  \label{tab:ablation}
  \vskip -10pt
\end{table}

\begin{figure*}[t!]
  \centering
\subfloat[AUC score on Wiki]{%
    \includegraphics[width=0.25\linewidth]{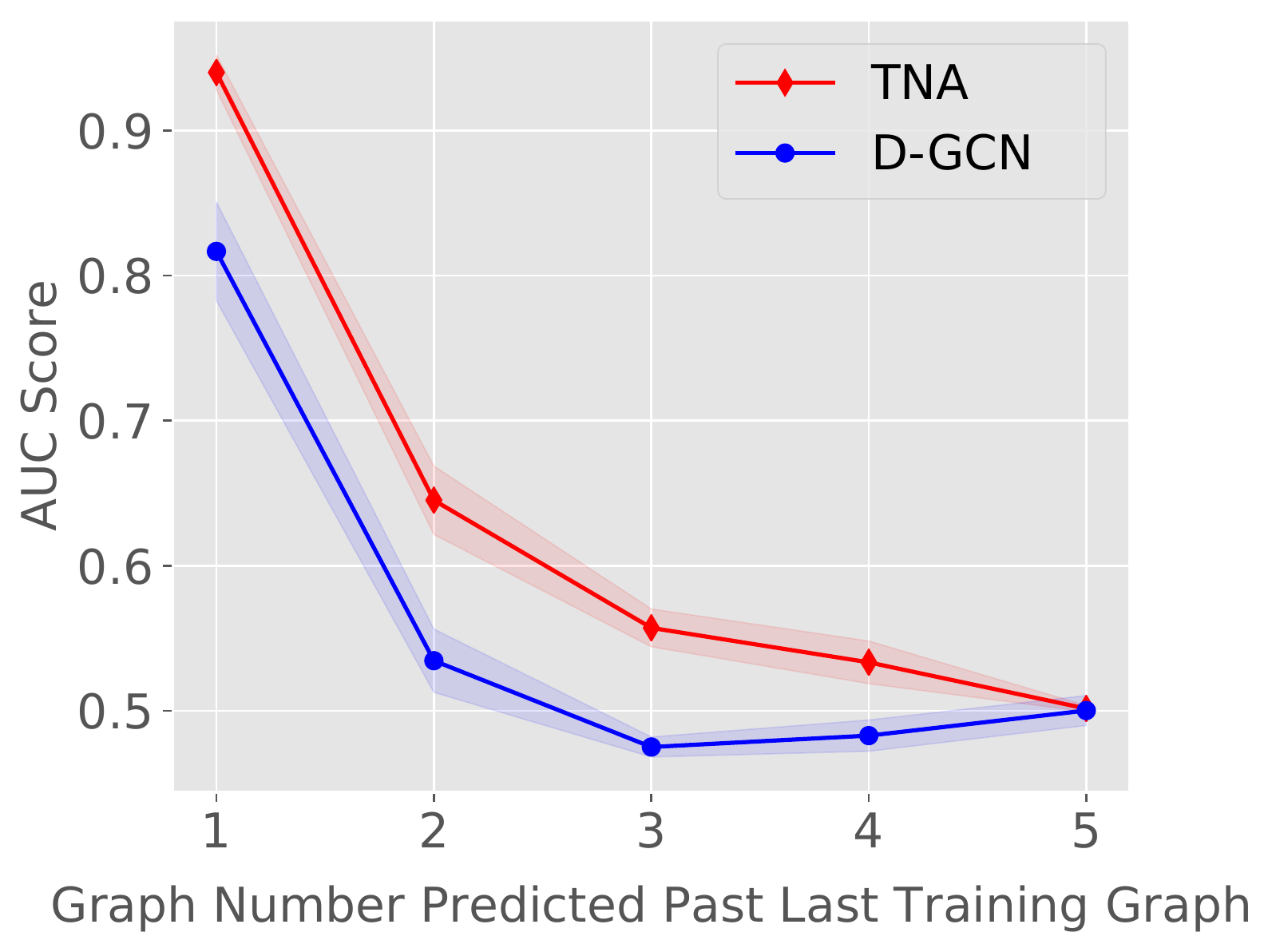}} 
  \label{PRmiCA5}\hfill
\subfloat[AP score on Wiki]{%
    \includegraphics[width=0.25\linewidth]{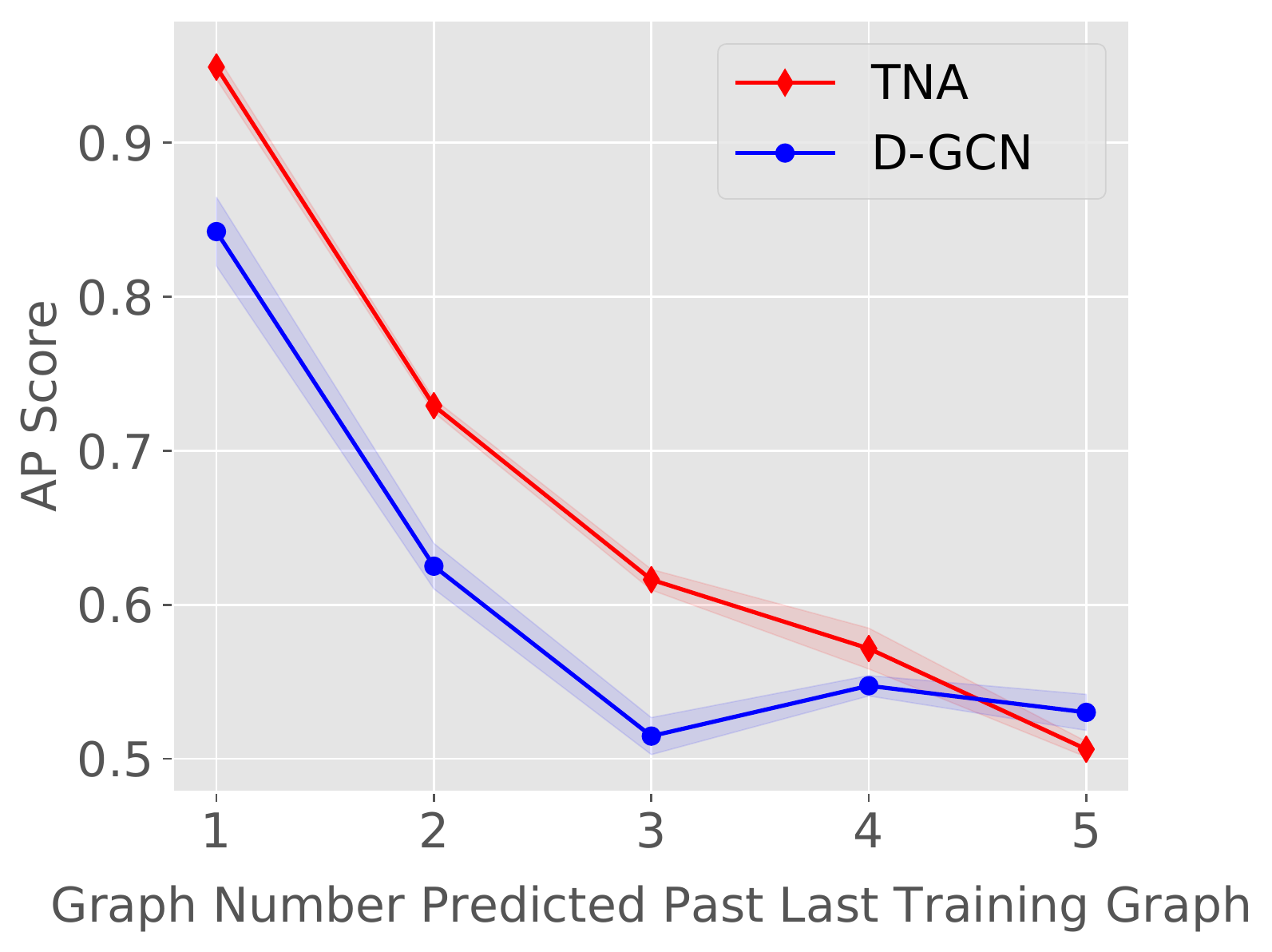}}
  \label{PRmiFB5}\hfill
\subfloat[AUC score on UCI]{%
    \includegraphics[width=0.25\linewidth]{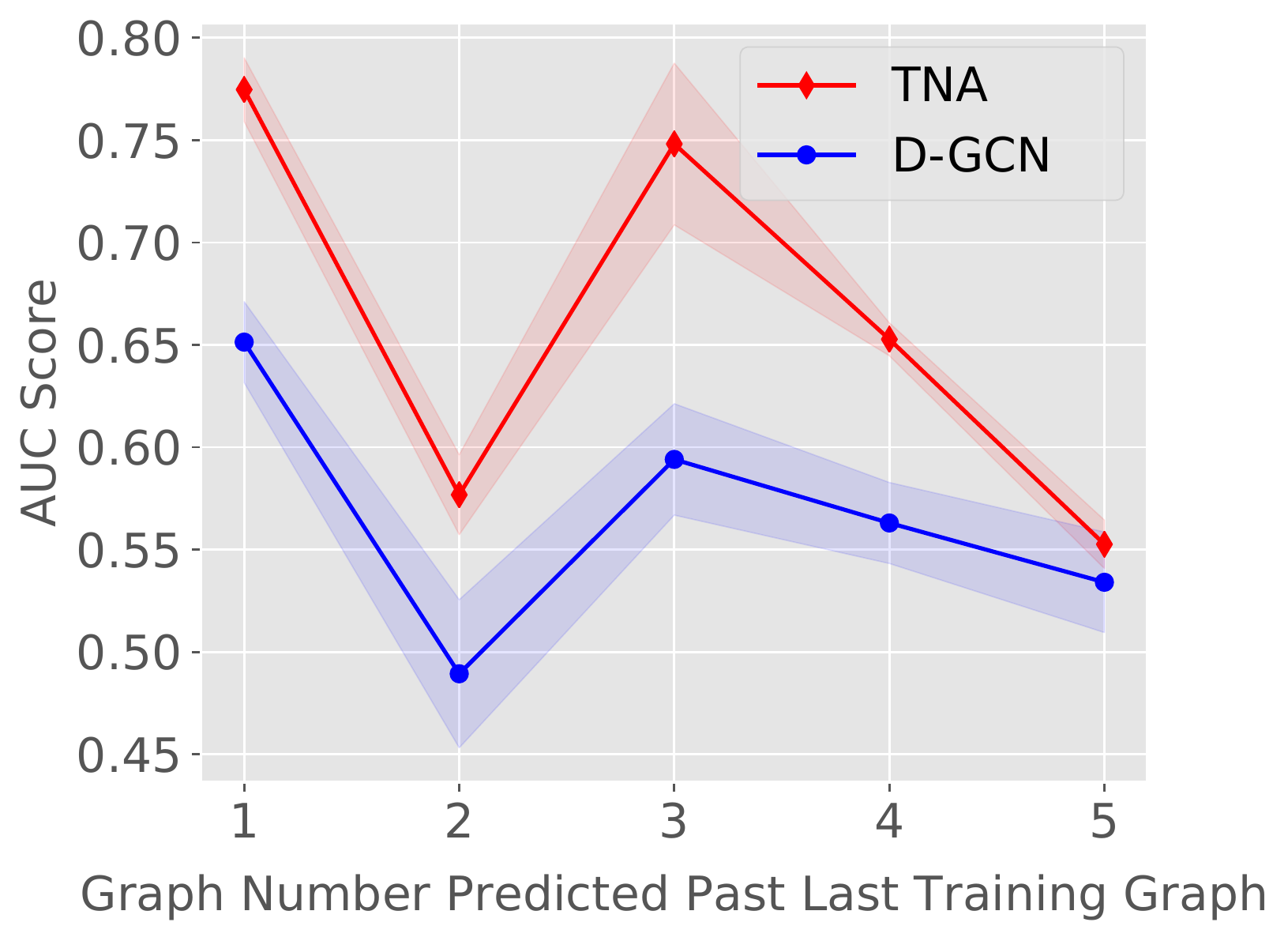}}
  \label{PRmiGN5}\hfill
\subfloat[AP score on UCI]{%
    \includegraphics[width=0.25\linewidth]{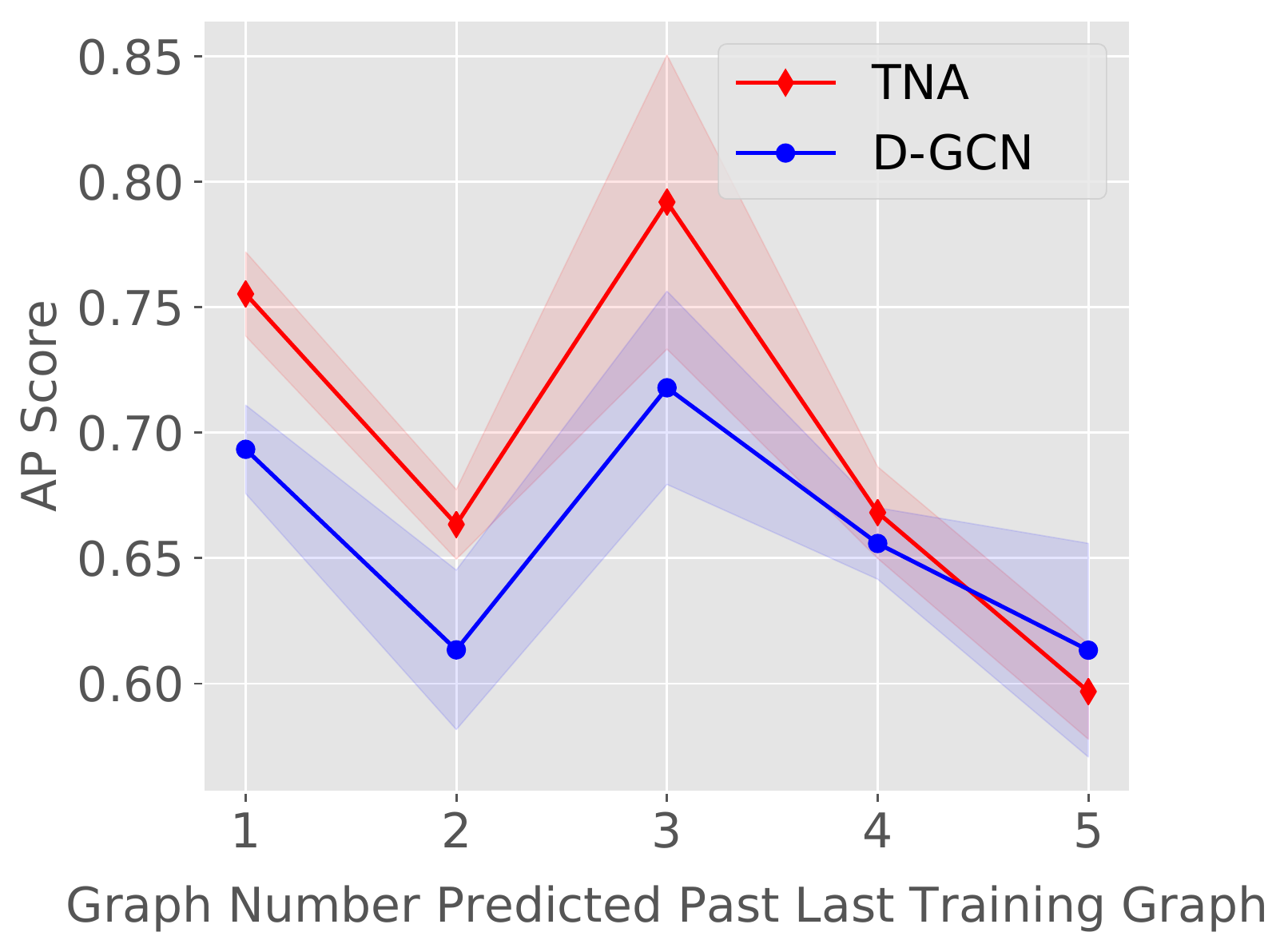}}
  \label{PRmiWI5}\\
\caption{AUC and AP for the Wiki and UCI datasets when predicting new edges $n$ number of time points away from the end of the training sequence. Results presented as the mean of three uniquely trained models, each with a different random seed.}
\label{fig:futre_graph_pred} 
\vskip -10pt
\end{figure*}

\begin{table}[h!]
  \centering
  \begin{tabular}{l l c c}
  \toprule
  \textbf{Dataset}  & \textbf{Approach} & \textbf{AUC} & \textbf{AP}\T\B \\
  \midrule \midrule
\multirow{9}{*}{SBM} & GAE   & $\bm{0.505\pm0.018}$ & $0.451\pm0.009$  \\
                    & GVAE   & $0.500\pm0.012$   & $\bm{0.503\pm0.011}$   \\
                    & TO-GAE   & $0.504\pm0.017$  & $0.451\pm0.008$  \\
                    & TO-GVAE   & $0.500\pm0.012$  & $0.503\pm0.011$  \\
                    & DynAE   & $0.023\pm0.003$   & $0.431\pm0.008$  \\
                    & DynRNN    & $0.039\pm0.005$   & $0.348\pm0.009$  \\
                    & DynAERNN    & $0.008\pm0.000$   & $0.308\pm0.000$  \\
                    & D-GCN    & $0.458\pm0.017$   & $0.458\pm0.017$  \\
                    
                    \cline{2-4}
                    & \textbf{TNA} & $0.502\pm0.024$  & $0.502\pm0.017$\T  \\

\midrule

\multirow{10}{*}{R-Cora} & GAE  & $0.501\pm0.015$   & $0.500\pm0.0100$  \\
                      & GVAE   & $0.491\pm0.011$   & $0.494\pm0.002$   \\
                      & TO-GAE   & $0.500\pm0.013$   & $\bm{0.502\pm0.009}$ \\
                      & TO-GVAE   & $0.490\pm0.011$   & $0.494\pm0.011$ \\
                      & DynAE   & $0.356\pm0.001$   & $0.479\pm0.003$  \\
                      & DynRNN   & $0.308\pm0.011$   & $0.381\pm0.011$   \\
                      & DynAERNN    & $0.201\pm0.000$   & $0.346\pm0.000$   \\
                      & D-GCN   & $\bm{0.502\pm0.011}$  & $0.500\pm0.008$   \\    

                      \cline{2-4}
                      & \textbf{TNA} & $0.493\pm0.012$ & $0.493\pm0.012$\T  \\
  \bottomrule
  \end{tabular}
  \caption{Next graph prediction results on sythnetic graphs presented as mean values with standard deviation when predicting at each point in the time series.}
  \label{tab:nextgraph-synth}
  \vskip -10pt
\end{table}

\subsection{Next Graph Link Prediction}
\label{sec:lp}

As the main focus of our model, we present results for predicting new edges in the next temporal graph, using the procedure detailed in Algorithm \ref{algo:training_procedure}, in Table \ref{tab:lp}\footnote{DynRNN is missing for the Wiki dataset as it could not fit in GPU memory.}. The table shows that TNA significantly outperforms the baseline approaches when predicting new edges in the next graph at all points along the time series. Compared with the Dyn* family of approaches, it is striking to note the significant number of parameters required by the models (often well over an order of magnitude more) and their poor performance in predicting new edges. We believe it is highly likely that this family of models is using the extra parameters to over-fit to the edges that do not change over time, resulting in bad predictive capability for the ones that do. It is also interesting to note that, compared with the D-GCN approach, TNA is better able to capture the dependences needed for good long-term prediction. For two datasets our model improves the past graph evolution data it has to learn from. This is demonstrated by the increasing AUC and AP scores for the Bitcoina and UCI datasets. However, all approaches struggle on the synthetic datasets due to the inherent random nature, as seen in Table \ref{tab:nextgraph-synth}. 

\subsection{Full Graph Reconstruction}
\label{sec:full_graph_rec}

To measure the ability of the representations learned by the TNA model to be used as general purpose embeddings, we look at the problem of future graph reconstruction. Here, the performance of the model at predicting the presence of edges in the full graph $G_t$ (given $G_1 .. G_{t-1}$) is measured -- highlighting how we do not sacrifice performance at predicting existing edges. This will allow us to investigate the ability of the model to predict not only new edges, but that existing edges have not been removed. As before, a new model is trained to predict the final graph in the sequence given all previous time points, with the final results presented as the mean over all graphs in the sequence. However, instead of predicting edges which have appeared since the last time point, here the results are for a balanced set of random sampled positive and negative edges in $E_t$ which may or may not include ones formed since the previous time point.

\begin{table}[h!]
  \centering
  \begin{tabular}{l l c c}
  \toprule
  \textbf{Dataset}  & \textbf{Approach} & \textbf{AUC} & \textbf{AP}\T\B \\
  \midrule \midrule
\multirow{5}{*}{Bitcoina} & DynAE   & $0.830\pm0.068$   & $0.844\pm0.050$  \\
                    & DynRNN    & $0.922\pm0.059$   & $0.937\pm0.039$  \\
                    & DynAERNN    & $\bm{0.968\pm0.057}$   & $\bm{0.981\pm0.034}$  \\
                    & D-GCN   & $0.919\pm0.021$   & $0.934\pm0.016$  \\
                    
                    \cline{2-4}
                    & \textbf{TNA} & $0.932\pm0.024$  & $0.945\pm0.018$ \T  \\

\midrule

\multirow{5}{*}{UCI} & DynAE   & $0.905\pm0.061$   & $0.908\pm0.055$  \\
                      & DynRNN   & $0.957\pm0.015$   & $0.954\pm0.010$   \\
                      & DynAERNN    & $\bm{0.988\pm0.014}$   & $\bm{0.993\pm0.009}$   \\
                      & D-GCN   & $0.829\pm0.019$  & $0.862\pm0.014$   \\    

                      \cline{2-4}
                      & \textbf{TNA} & $0.821\pm0.015$ & $0.847\pm0.012$\T  \\

  \midrule

  \multirow{4}{*}{Wiki} & DynAE   & $0.765\pm0.088$   & $0.795\pm0.062$  \\
                        & DynAERNN    & $0.882\pm0.072$   & $0.934\pm0.037$   \\
                        & D-GCN   & $0.905\pm0.019$  & $0.936\pm0.015$   \\    
  
                        \cline{2-4}
                        & \textbf{TNA} & $\bm{0.919\pm0.014}$ & $\bm{0.945\pm0.007}$\T  \\
  \bottomrule
  \end{tabular}
  \caption{Results for predicting both new and old edges in the final graph in the sequence, presented as a mean and standard deviation over the whole time sequence. A bold value indicates the highest score for that metric. TNA remains competitive with, and even beats many baseline approaches with a much greater number of parameters.}
  \label{tab:fullgraph}
  \vskip -10pt
\end{table}

The results for this experiment are presented in Table \ref{tab:fullgraph} where for the sake of brevity, we compare with only the temporal baselines. It is obvious that many of the baselines, especially the Dyn* family of approaches perform much better at predicting existing edges than new ones. This further suggests that they are utilising their larger set of parameters to, in some way, over-fit to edges which have been in the graph for a longer length of time, which form the vast majority. However despite this, our TNA approach still performs well at this task, displaying comparable performance with the baseline approaches and even outperforming them on the Wiki dataset. This further strengthens the argument that having recurrence at each hop in the neighbourhood aggregation produces a better representation, whilst requiring fewer parameters.

\subsection{Future Graph Evolution}
\label{sec:future_graph_evo}

For our final experiment, we investigate how TNA performs when predicting new edges further into the future than the next graph. We train the models on 70\% of the available temporal history, then predict new edges and compare with the remaining ground truth data. To achieve this, we feed the graph predicted by the models as the next graph in the sequence back into the model, which is subsequently used to predict the next graph. This is similar to using RNNs as generative models to produce text data \cite{sutskever2011generating} and can be seen as a combination of both the previous tasks. Figure \ref{fig:futre_graph_pred} displays the results for this task, where we compare with the closet baseline from Section \ref{sec:lp}. The results show how TNA is better able to predict new edges into the future, emphasising its capability to learn a good temporal representation for the vertices.

\section{Conclusion}
\label{sec:conclusion}

Many real-world graph datasets have rich and complex temporal information available which is disregard by the majority of the current approaches for creating vertex representations. In this paper, we have introduced the Temporal Neighbourhood Aggregation model for representation learning on large, complex temporal graphs. Our approach demonstrates excellent performance through extensive experimental evaluation, beating several competing temporal and static models, when predicting future edges not seen in the training data. The TNA model can learn complex temporal patterns present at multiple depths within a vertices neighbourhood, creating the final vertex representation via the use of variational sampling. 

For future work, we will investigate replacing the GCN in our model with an approach designed for inductive learning \cite{hamilton2017inductive} to allow for training on even larger graph datasets, as well as enabling vertex arrival to be modelled. We also plan to experiment using the learned representations for additional tasks, such as temporal classification.

\section*{Acknowledgement}

We gratefully acknowledge the support of NVIDIA Corporation with the donation of the GPU used for this research. Additionally we thank the Engineering and Physical Sciences Research Council UK (EPSRC) for funding.

\bibliographystyle{IEEEtran}
\bibliography{paper_ref}

\end{document}